# Multibridge VO$_2$-Based Resistive Switching Devices in a Two-Terminal Configuration

*Xing Gao, Thijs J. Roskamp, Timm Swoboda, Carlos M. M. Rosário, Sander Smink, Miguel Muñoz Rojo, Hans Hilgenkamp\**

X. Gao, T. J. Roskamp, C. M. M. Rosário, S. Smink, H. Hilgenkamp
Faculty of Science and Technology and MESA+ Institute for Nanotechnology, University of Twente, 7500 AE Enschede, The Netherlands
E-mail: h.hilgenkamp@utwente.nl

T. Swoboda, M. Muñoz Rojo
Department of Thermal and Fluid Engineering, Faculty of Engineering Technology, University of Twente, 7500 AE Enschede, The Netherlands

M. Muñoz Rojo
Instituto de Micro y Nanotecnología, IMN-CNM, CSIC (CEI UAM+CSIC), 28760 Tres Cantos Madrid, Spain

Keywords: vanadium dioxide, resistive switching, multilevel operation, electrically driven insulator-to-metal transition, thermal crosstalk, scanning thermal microscopy

Vanadium dioxide (VO$_2$) exhibits a hysteretic insulator-to-metal transition near room temperature, forming the foundation for various forms of resistive switching devices. Usually, these are realized in the form of two-terminal bridge-like structures. We show here that by incorporating multiple, parallel VO$_2$ bridges in a single two-terminal device, a wider range of possible characteristics can be obtained, including a manifold of addressable resistance states. Different device configurations are studied, in which the number of bridges, the bridge dimensions and the interbridge distances are varied. The switching characteristics of the multibridge devices are influenced by the thermal crosstalk between the bridges. Scanning Thermal Microscopy has been used to image the current distributions at various voltage/current bias conditions. This work presents a route to realize devices exhibiting highly non-linear, multistate current-voltage characteristics, with potential applications in e.g., tunable electronic components and novel, neuromorphic information processing circuitry.

## 1. Introduction

To overcome limitations of computing systems based on von Neumann architectures, like excessive data transfer between memory and logic units, neuromorphic device concepts that mimic the function of neurons and synapses are of particular interest.[1,2] The neuromorphic computing circuitry requires novel circuit elements with tunable resistance states, non-linear response functions and, for the case of spiking neuromorphic circuitry, adaptable dynamic behavior.[1,3,4] $VO_2$ is an attractive candidate material to fulfill several of these roles, because it exhibits a near-room-temperature, hysteretic insulator-to-metal transition (IMT) with resistivity changes of several orders of magnitude.[5–10] The IMT can be tuned by chemical doping,[11] epitaxial strain,[12,13] and external stimuli.[14–16] Particularly, it can be triggered by electrical voltage/current and the associated Joule heating.[16–21] The electrical and thermal conductivity of $VO_2$ are highly temperature-dependent, leading to nonlinear dynamics in an electrothermal feedback loop.[22,23] Using such nonlinear behavior, Yi et al. have demonstrated a range of neuronic spiking patterns in $VO_2$-based neuromorphic circuits.[24] Furthermore, tunable multilevel resistive states have been achieved in single $VO_2$ bridges, with the outlook that devices with multiple parallel bridges may provide a higher degree of control.[20]

Here, we extend our studies on two-terminal $VO_2$ devices incorporating such multiple parallel bridges. To characterize the switching behavior, we perform both voltage-controlled and current-controlled measurements, which result in different current-voltage (*I-V*) characteristics as will be elucidated below. We have employed Scanning Thermal Microscopy (SThM) to directly image the current distribution in the devices.[25] SThM uses a special thermo-resistive probe with high thermal sensitivity (< 1 K) that enables the characterization of thermal phenomena on the sample surface with nanoscale spatial resolution.[26–28] For our $VO_2$ bridges, the local heating generated is directly linked to the current flow, due to Joule heating.[17,29]

## 2. $VO_2$-based double-bridge devices

In **Figure 1**a, the device configuration is sketched. The most basic version, previously reported,[20] consists of a single $VO_2$ bridge contacted in a two-terminal configuration. In the same figure, a double-bridge device is shown, in which the two-terminal configuration is maintained. However, this structure consists of two parallel bridges with length *L*, width *W* and interbridge spacing *d*. Later on, configurations with more than two bridges will be discussed as well. Figure 1b shows the switching characteristics of a single bridge, in voltage- and current-sweep mode, respectively. These data are in line with our previous report,[20] whereby it is noted that all measurements discussed here are conducted after a first forming cycle. For a voltage-controlled sweep, the heating generated at the switching point (here at $V$ = 15 V) increases abruptly by the sharp rise of the current, up to the compliance current ($I_{CC}$), which results in a sharp increase in dissipated power. For a current-controlled sweep, on the other hand, the heating develops more gradually due to the associated voltage drop and can be stabilized. On the reset path, using the current bias, a region with a negative differential resistance (NDR) can be identified. Figure 1c shows the numerically computed differential conductance $dI/dV$, for a limited range of values. The supplementary figure **Figure S1** shows the full scale of the differential conductance, including the NDR values.

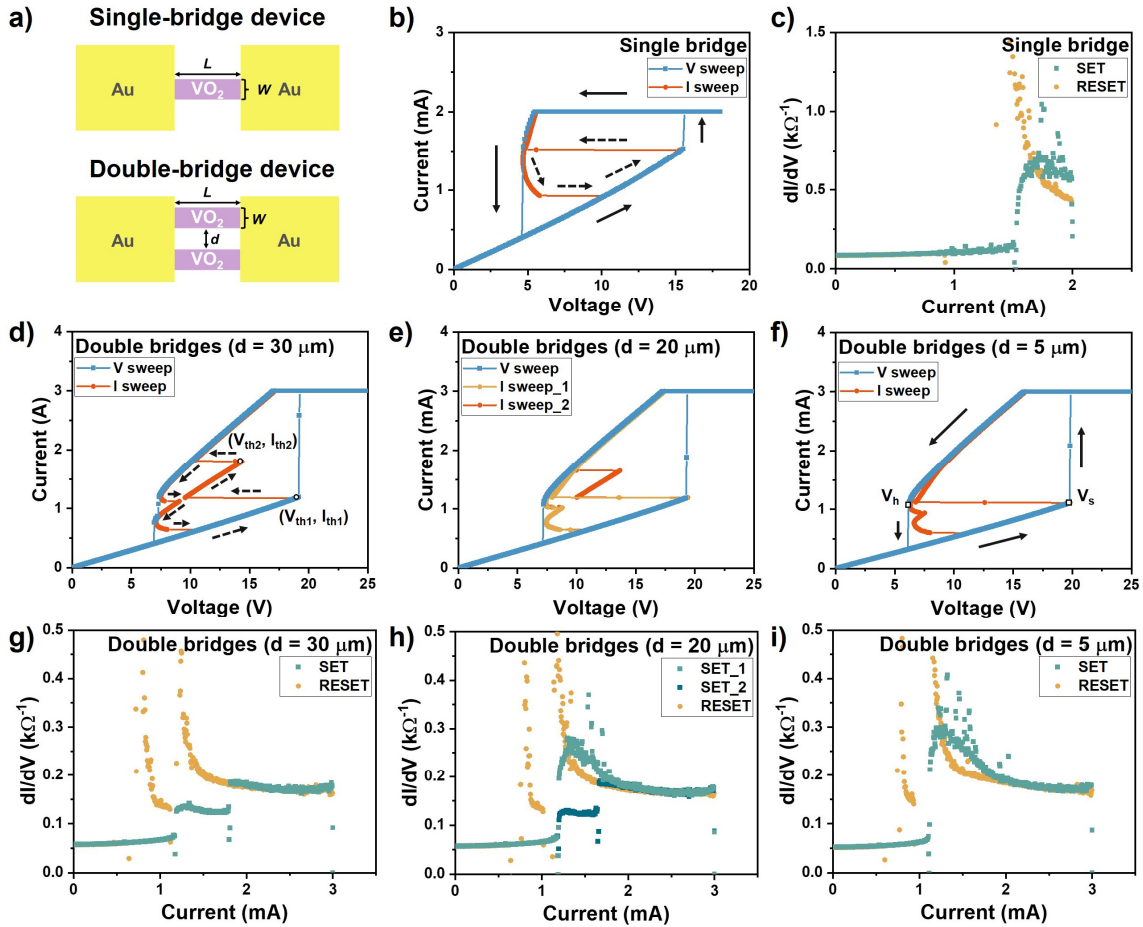

**Figure 1.** a) Schematic top-view of the VO$_2$-based single-bridge and double-bridge device. The width ($W$), length ($L$), and spacing ($d$) of the patterned VO$_2$ parallel bridges are as labeled. b) Voltage-controlled (direction indicated with solid arrows) and current controlled (direction indicated with dashed arrows) *I-V* characteristics and c) differential conductance ($dI/dV$) as a function of applied current during the set and reset process of a single-bridge device ($L$ = 20 µm, $W$ = 5 µm). $I_{CC}$ = 2 mA. Voltage-controlled and current-controlled *I-V* characteristics and of double-bridge devices ($L$ = 20 µm, $W$ = 5 µm) with different spacings: d) $d$ = 30 µm, e) $d$ = 20 µm, and f) $d$ = 5 µm. $I_{CC}$ = 3 mA. The terminology of threshold voltages ($V_{th}$) and currents

($I_{th}$) for current sweeps are indicated in d). The terminology of set voltage ($V_s$) and hold voltage ($V_h$) for voltage sweeps are indicated in f). Differential conductance $dI/dV$ during the set and reset process for corresponding devices: g) $d$ = 30 µm, h) $d$ = 20 µm, i) $d$ = 5 µm. The range is selected to clearly see the conductance steps. See Figure S1 for the full vertical axis range of $dI/dV$ values.

Figures 1d-f show measurements on a double bridge structure with a fixed length ($L$ = 20 µm) and width ($W$ = 5 µm) for different values of the spacing between the bridges, i.e., 30 µm, 20 µm and 5 µm. Here, the difference between the current- and voltage-bias sweeps becomes extra apparent. While with a voltage bias, we again see a single, big increase in the current at the switching point, for the current bias we see - depending on the separation - multiple switches, as can be expected from a sequential switching of the individual bridges.

In the double-bridge devices, the switching behaviors of the parallel bridge devices are determined by the intrinsic IMT of individual bridges and the thermal interaction between them. In this work, several available knobs at a device level are tuned for manipulation of the switching behaviors. Double-bridge devices, multiple-bridge devices, and multi-width bridge devices are discussed in the following sections. Despite the rising complexity, there are some basic trends that have been observed. One is that more potential switches and intermediate resistive states can be achieved by adding more bridges in parallel.

Also, the spacing between bridges influences the number of switching events and the power required for subsequent switches. This is clear from Figure 1d-f and **Figure S2**. If the bridges are far away from each other, for example $d$ = 30 µm (Figure 1d), the thermal crosstalk is small and the bridges switch individually and sequentially. In this case a stable intermediate resistive state occurs. If the bridges are close to each other, for example $d$ = 5 µm (Figure 1f), the heat dissipation of one bridge will affect the other greatly and both bridges will be triggered simultaneously, so that there is only one switching event during the set process. As shown in Figure 1e, there is a critical spacing value, in this experiment $d$ = 20 µm, where the device shows simultaneous switching or individual switching randomly.

From the current-controlled *I-V* characteristics, the differential conductance ($dI/dV$) of the measured devices is calculated and plotted as a function of applied current, see Figures 1g-i and Figure S1. The $dI/dV$ values near the switching point or in the NDR region diverge to large positive and negative values. When the device resistance settles, the $dI/dV$ should scale with the number of conductance paths available for the current.[30] Indeed, we see that when there is only one switch the change of differential conductance per switch is double to the case when there are two consecutive switching events.

Interestingly, there are several NDR regions in the current-controlled sweep, where the differential resistance ($dV/dI$) of the device is negative. NDR can generally be found in materials that form higher current density channels relative to the rest of the material under electrical stimuli.[31] There are two types of NDR, a 'snapback NDR' refers to a discontinuous transition from positive to negative differential resistance which is a generic response of materials with a strong temperature dependent conductivity, while an 'S-type NDR' refers to a smooth transition that can emerge from tuning the material conductivity, device area or ambient temperature.[22,32] Both types, and sometimes even more complex combinations of them, are observed in our devices. The dynamical instabilities associated with NDR are of great interest for potential applications, such as selectors, threshold switches, amplifiers, and oscillators.[22] With the thermally induced IMT in $VO_2$, the dynamics and instabilities result from an electrothermal feedback loop, as the electrical and thermal properties are highly temperature-dependent.[4] In our devices, the snapback NDR is usually observed during the set process due to the sharp transition from low conductivity to high conductivity, while the S-type NDR is observed during the reset where the $VO_2$ bridges are experiencing a more gradual transition.[18,33]

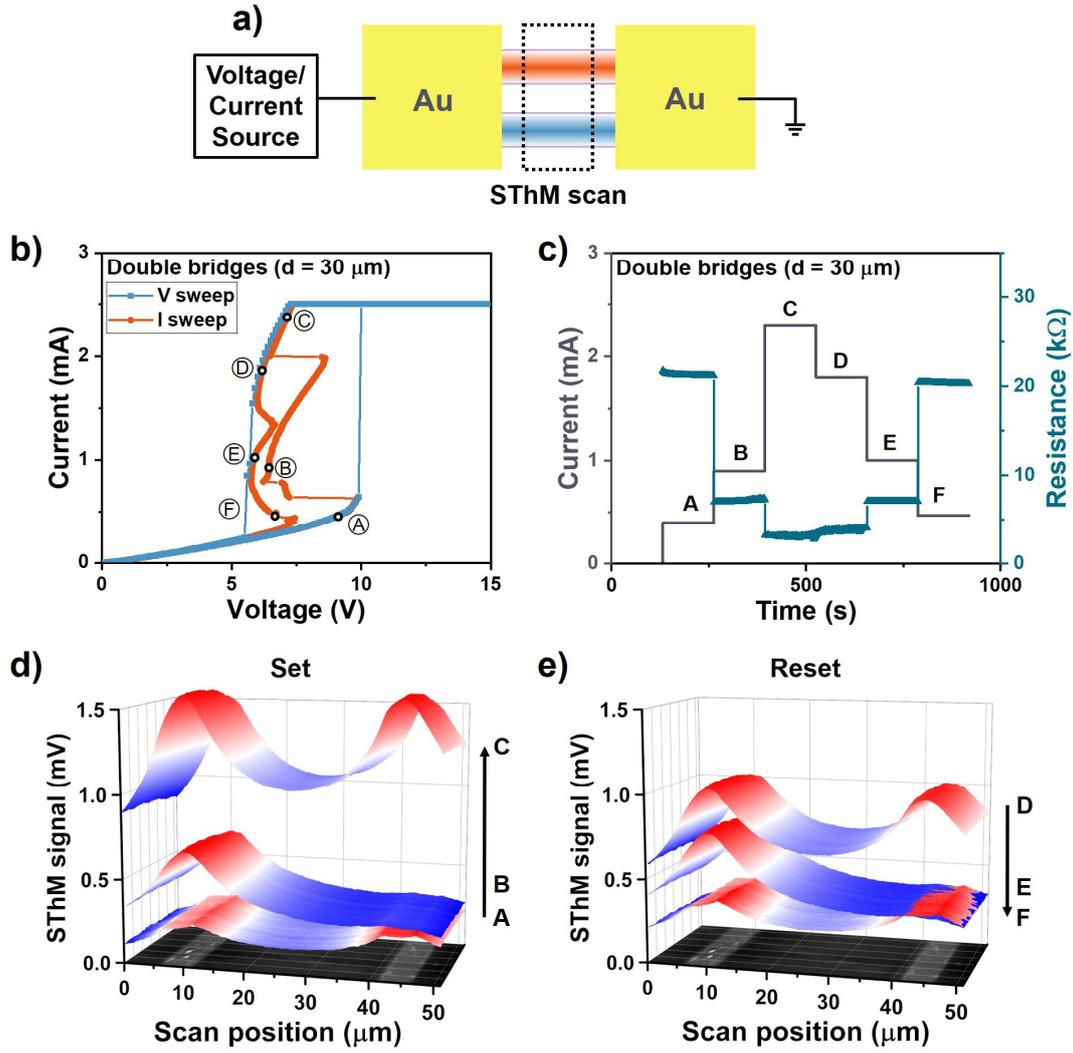

**Figure 2.** a) Schematic of the SThM measurement. The black dashed line box indicates the scan area. The color codes assigned to the bridges denote their status, with red indicating that the bridge is ON (i.e., low resistance) while blue represents a bridge that is OFF (i.e., high resistance). b) Voltage-controlled and current-controlled *I-V* characteristics of a VO$_2$-based double-bridge device ($L$ = 20 μm, $W$ = 5 μm, $d$ = 30 μm). $I_{CC}$ = 2.5 mA. The points where SThM measurements were taken are labeled as follows; A: 0.4 mA, B: 0.9 mA, C: 2.3 mA, D: 1.8 mA, E: 1 mA and F: 0.47 mA. c) Applied current and measured resistance (*V/I*) during SThM measurements. d) and e) Qualitative 3D SThM thermal maps during the set and reset process, respectively. The points where the scans were taken are indicated beside their respective maps and the arrow shows the order of the scans. The 2D surface topography images at the bottom are obtained using the SThM tip in Atomic Force Microscopy mode and are shown as a guide for the position of the bridges in the device.

Assuming that the voltage is equal across both bridges and decreases linearly along the device, the current can be determined from the Joule heating. Therefore, the current distribution and switching behavior in our multi-bridge devices can be visualized by mapping the Joule heating locally using SThM, as is shown in **Figure 2**a. Figure 2b shows that the bridges are switched individually under the applied current bias for the well-separated double-bridge device under study here. During the SThM scans, the device is set to stable intermediate resistive states by biasing at constant current values as shown in Figure 2c. The qualitative thermal maps obtained during the set process and the reset process maintaining constant measurement conditions are plotted in Figure 2d and Figure 2e, respectively. In this, the SThM signal correlates with the temperature increase of the bridges. By increasing the bias current from zero to a value below the threshold current for switching $I_{th}$, the current distributes evenly over the two bridges (scan A). When one of the bridges switches, the heating is redistributed indicating that almost all the current flows through the bridge that has switched ON (scan B). By further increasing the bias current, there will be an increase in the current flowing through both bridges according to their differences in resistance and, potentially supported by heat transferred from the bridge with the highest current, also the remaining bridge switches to a low-resistance state (scan C).

The reset process is basically the inverse process of the set behavior, but follows a different trace, due to the hysteretic resistance versus temperature characteristics of the $VO_2$. Starting with two bridges in the low-resistance state (scan D), the right bridge is switched OFF when the current decreases (scan E), and finally the remaining bridge is also switched to the high-resistance state and the current is almost evenly distributed over both bridges again (scan F). While in principle the switching sequence between the two bridges for the set and reset process could be different, we see in Figure 2 as well as in all our other measurements thus far, that the order of switching back to the high-resistance state is the reverse of the switching events to the low-resistance state. Possible contributors for the fixed switching order are the intrinsic differences between the two bridges, including slight variations in geometry, dimensions, material composition, and local defects. The effects may have been amplified during the initial forming step, which could have defined the overall behavior of the bridges.

In the supplementary information (Figure S2) the SThM characteristics of the well-separated bridges are compared with the case for bridges in close proximity. In accordance with our observation of a single simultaneous switching of both bridges in the latter case, we see only one wide heating path covering both bridges in the SThM measurements for those devices.

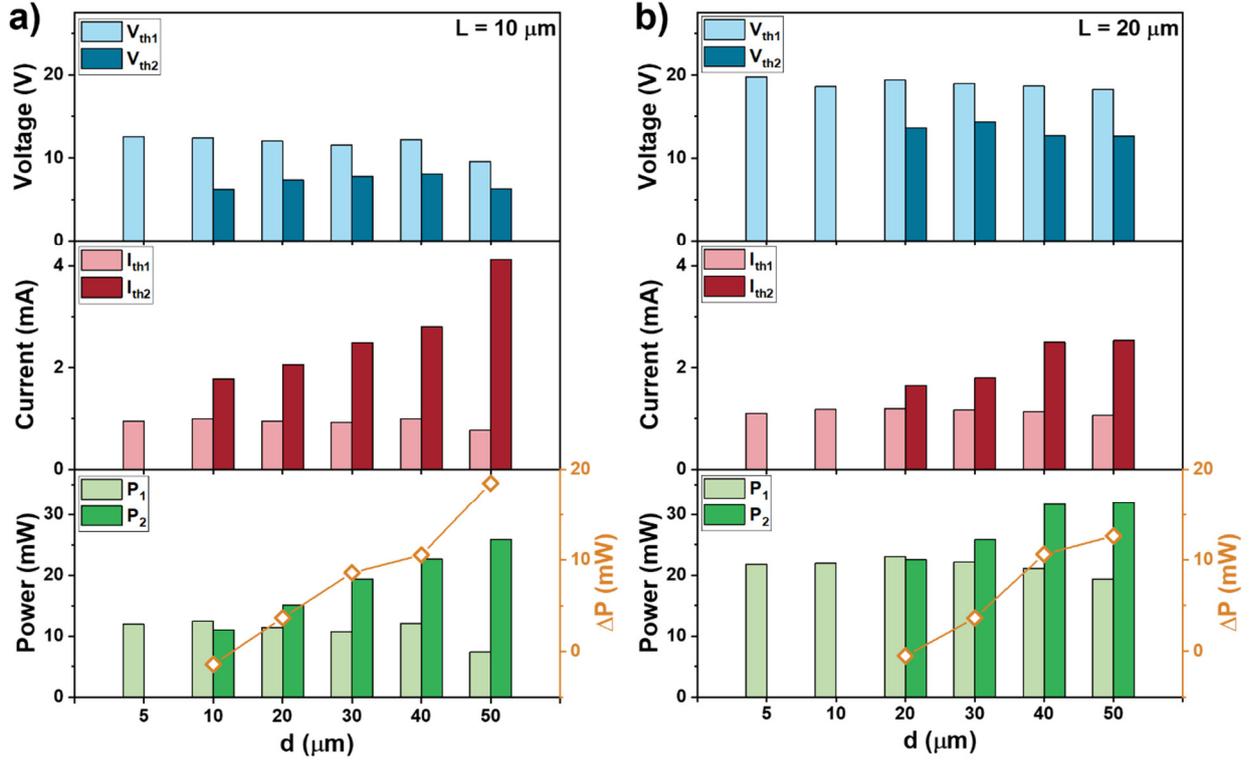

**Figure 3.** Switching parameters of current-controlled measurements as a function of the bridge spacing for VO$_2$-based double-bridge devices: a) $L = 10$ μm, $W = 5$ μm and b) $L = 20$ μm, $W = 5$ μm. The top panels show the threshold voltage ($V_{th}$). The middle panels show the threshold current ($I_{th}$). The bottom panels show the calculated switching power ($P_{th}$) and the power difference ($\Delta P$) between the first and the second snapback. Subscripts 1 and 2 stand for the first snapback and the second snapback, respectively.

**Figures 3**a and 3b show the correlation between the switching parameters and the bridge spacing $d$ for double-bridge devices with different bridge length, $L = 10$ μm and $L = 20$ μm, respectively. The threshold voltage $V_{th}$ and power $P_{th}$ in these current-controlled experiments both scale with bridge length, while the threshold current $I_{th}$ does not. The shorter bridges can be switched individually at a smaller spacing compared to the longer bridges, which is beneficial for down scaling the device further. The power needed for the first snapback ($P_1$) is found to be not strongly dependent on the spacing while the power required for the second one ($P_2$) is correlated with the spacing. At the critical point, $d = 10$ μm in Figure 3a and $d = 20$ μm in Figure 3b, the first switching power is already enough for both bridges to be switched. However, due to fluctuation in the thermal crosstalk, there is also potential individual switching within the device and sometimes it dominates the switching behavior. The power difference

($\Delta P$) between $P_2$ and $P_1$ scales with the bridge spacing, in line with the consideration that more power is required for the secondary switch when the thermal crosstalk becomes less significant.

Temperature-dependent measurements are also performed on double-bridge devices (see **Figure S3** and **Figure S4**). With increasing operating temperature, $V_s$, $V_h$ and the hysteresis window between them for the voltage-controlled measurements reduce. For the current-controlled measurements of the device with large spacing ($d = 30$ μm), $I_{th}$, $P$ and $\Delta P$ also decrease with increasing temperature. Interestingly, the device with smaller spacing ($d = 5$ μm) starts to show individual switching when temperature increases. The critical temperature is 300 K where the device shows a secondary switch after the NDR region. When operating the device closer to the transition temperature, the required power for switching becomes lower, therefore the thermal cross talk also decreases.

### 3. VO$_2$-Based Multiple-Bridge Devices

Based on the systematic studies on double-bridge devices, we further fabricated triple-bridge devices (**Figure 4**a). The addition of an extra bridge introduces the potential for more switches and resistive states. The current-sweep curves in Figures 4b and 4c show there are three distinct snapbacks and a maximum of four stable resistive states for the triple-bridge devices. To image the switching dynamics, SThM thermal maps of each resistive state are measured (**Figure S5**b). It is observed that, in this particular measurement, the bridges are switched in sequence from right to left as the applied current increases, corresponding to the individual snapbacks in the *I-V* curves.

To obtain even more resistance states, two more bridges are added to the device design, thus forming a quintuple-bridge device (Figure 4d). As shown in Figures 4e and 4f, the current-voltage characteristics of these devices show five distinct snapbacks and a maximum of six stable resistive states, all in a single two-terminal device without additional external stimuli. The SThM thermal maps in Figure S5d show that the bridges are switched individually, in this case from the bridge in the middle to the ones at the edges. To be noted, for both triple-bridge and quintuple-bridge devices, subsequent switches predominantly happen to the neighboring bridge of the ON-state bridge for $L > d$. For $L <\sim d$, however, stochasticity can be induced as will be further discussed. This observation provides further proof that thermal crosstalk is the main influential factor for the sequential switching process.

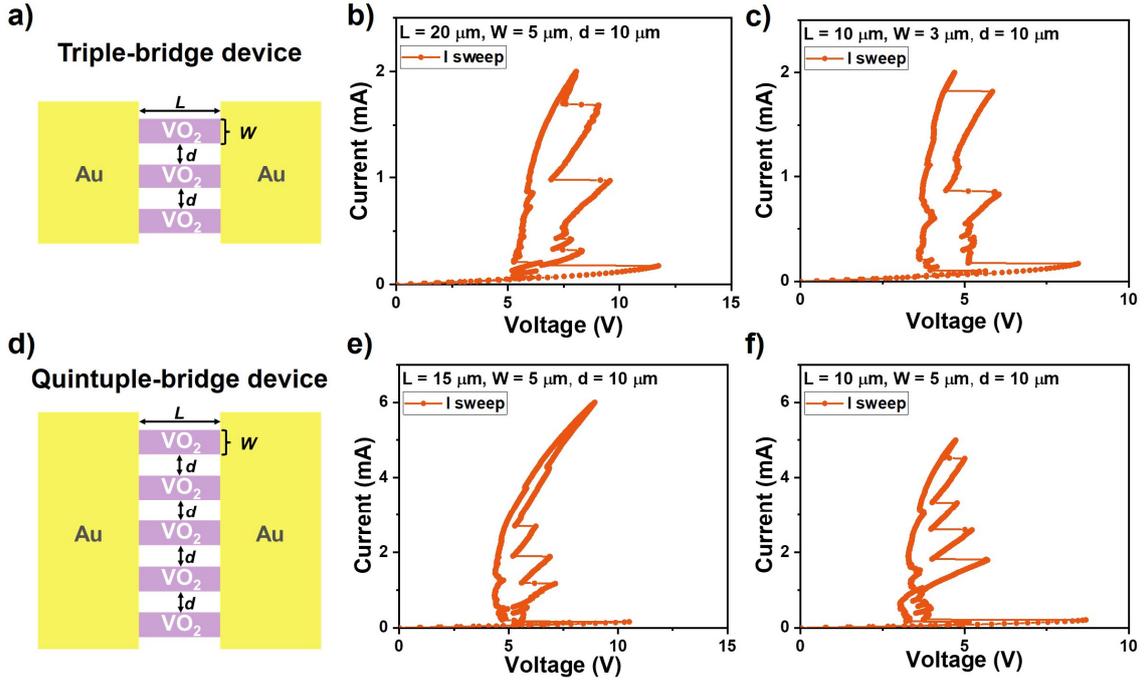

**Figure 4.** a) Schematic of the fabricated VO$_2$-based triple-bridge device (top view). Current-controlled *I-V* characteristics of VO$_2$-based triple-bridge devices: b) $L = 20$ μm, $W = 5$ μm, $d = 10$ μm and c) $L = 10$ μm, $W = 3$ μm, $d = 10$ μm. d) Schematic of the fabricated VO$_2$-based quintuple-bridge device (top view). Current-controlled *I-V* characteristics of two VO$_2$-based quintuple-bridge devices: e) $L = 15$ μm, $W = 5$ μm, $d = 10$ μm and f) $L = 10$ μm, $W = 5$ μm, $d = 10$ μm.

More resistive states can be achieved by adding even more switching elements in the device and the switching principles are still maintained. However, the increasing number of VO$_2$ bridges also introduces more complexity and stochasticity. As can be seen in the *I-V* characteristics in Figure 4, before the first stable intermediate state is achieved, the competition among bridges leads to irregular patterns in the low current range. Moreover, as revealed by the SThM scans of the multiple-bridge devices (**Figure S6** and **Figure S7**), the switching behavior after the first switch can also become complicated. For example, although the triple-bridge device in Figure 4c shows three distinct snapbacks, the order of switching bridges is completely different from the one in Figure 4b. As shown in Figure S6b and Figure S6c, the right bridge is set first (scan A and B), then for the second snapback, the other two OFF-state bridges are switched ON together while the right bridge is switched OFF, surprisingly (scan C). Finally, the right bridge is switched ON again creating the third snapback (scan at 2.2 mA). Even more complex phenomena are observed in the quintuple-bridge device in Figure 4f. The SThM thermal maps in Figure S7b show that the second bridge from the left is set first (scan A), but after the device reaches a stable resistive state, it loses the competition to the fifth bridge (scan B). Then the fourth bridge is triggered due to the thermal assistance of the ON-state fifth bridge (scan C). However, for the third snapback, the fourth bridge is switched OFF whereas the second bridge is switched ON again at the same time (scan D). Subsequently, the bridges in between the second and fifth ones are triggered with the same current (scan E) as they receive nearly equal heating from their neighbors. Finally, the last remaining bridge on the left is also switched ON with increasing current. To investigate the repeatability of the behavior in the low current range, an additional round of SThM scans is performed after scan F. This time the fifth bridge is set first without competing with the second one (scan A*) and the scans at higher currents remain the same. The first switched bridge can be different for different cycles.

As discussed above, due to the additional complexity, the intrinsic switching behavior is not always as straightforward as the current-controlled *I-V* characteristic would indicate. The competition among the nominally identical bridges for the first switch makes the set process in the low current range chaotic, leading to the unpredictable first switched bridge. Furthermore, more than one bridge can be switched ON simultaneously, sometimes accompanied by an ON-state bridge switching back to the OFF state. The surprising observation of an apparently stochastic order of switching for $L <\sim d$ cannot be explained fully by thermal crosstalk between steady-state bridges. We theorize that local thermal fluctuations could drive the observed stochasticity, which will be the subject of further research. This complexity may be further optimized and utilized in applications, for example in finite-state machine devices.[34]

## 4. VO$_2$-Based Multi-Width Bridge Devices

In the parallel-bridge devices, there are several variables to manipulate the switching behaviors, including the intrinsic IMT properties of the VO$_2$ bridges, the configuration of the devices, and the operating temperature. There is also an interplay between these factors, as was shown above. The intrinsic IMT determines the required switching power and the resistive states of the device. However, by tuning the operating temperature, the switching characteristic and critical spacing for synchronized/sequential/stochastic switching in multi-bridge devices can be changed. In the following, we will show an example of an additional degree of freedom, namely the geometry of the bridge.

So far, all the bridges in one device are identical with the same length, width and spacing. However, as mentioned in a previous section, competition between identical bridges can lead to unpredictable inherent switching behavior. We expect the geometry of bridges to strongly influence the switching behavior, as it affects the initial resistance, the switching power, and the generated heating of the bridges. The measurement results of two triple-bridge devices with multi-width bridges are shown in **Figure 5** and **Figure 6**. The devices contain bridges with the same length and spacing, but different widths and order. The wider bridge is placed at the edge of Device I (Figure 5a), and in the middle of Device II (Figure 6a). In current-controlled measurements, they both show three snapbacks and four resistive states, although the trend of the curves differs. The third switch of Device I (Figure 5b) is more intense and occurs at a lower current compared to Device II (Figure 6b). Furthermore, the *dI/dV* of the set process in Figure 5c indicates that there are three available current paths in Device I. The first two steps are of equal height while the third step is twice as high, suggesting the two 5 μm-wide bridges are switched first and followed by the 10 μm-wide bridge. It is confirmed by the SThM thermal maps in Figure 5d that, indeed, from the narrow one on the left to the wide one on the right the bridges are switched in sequence. In contrast, the *dI/dV* plots of Device II in Figure 6c are more scattered and there are only two steps with different height. The SThM measurements reveal its switching dynamics. As shown in Figure 6d, the left 5 μm-wide bridge is triggered first (scan A) and maintained ON with increasing current until the device reaches a steady resistive state (scan B). However, for the second snapback, the left bridge is switched OFF and at the same time the 10 μm-wide one in the middle is triggered (scan C). Increasing the current further, the left 5 μm-wide bridge is switched ON again while the right one remains OFF (scan D).

The behavior of measured multi-width bridge devices shows several recurring aspects. The narrow bridges in the devices tend to win the competition in the low current range and can be triggered first. This is because they exhibit more confined Joule heating due to their small dimensions and larger resistance. Moreover, once the wide bridges are switched ON, they dominate most of the current flowing through the device due to their small resistance in the low-resistance state. This can also lead to narrow bridges switching back to the OFF state. If one wants to avoid this, it is wise to place the wider bridge at the edge, like in Device I.

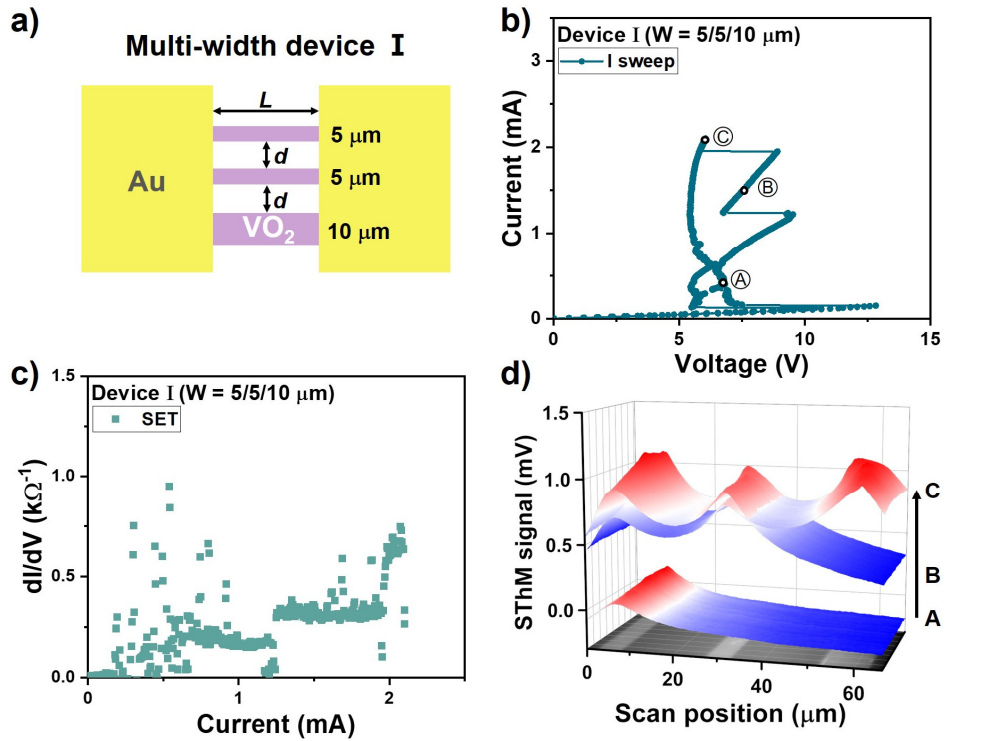

**Figure 5.** a) Schematic of Device Ⅰ (*L* = 20 μm, *W* = 5/5/10 μm, *d* = 20 μm) (top view). b) Current-controlled *I-V* characteristics and c) *dI/dV* plots of Device Ⅰ. d) 3D SThM thermal maps during the set process for Device Ⅰ. The points where SThM measurements were taken are labeled by letters (A: 0.4 mA, B: 1.5 mA, C: 2.2 mA).

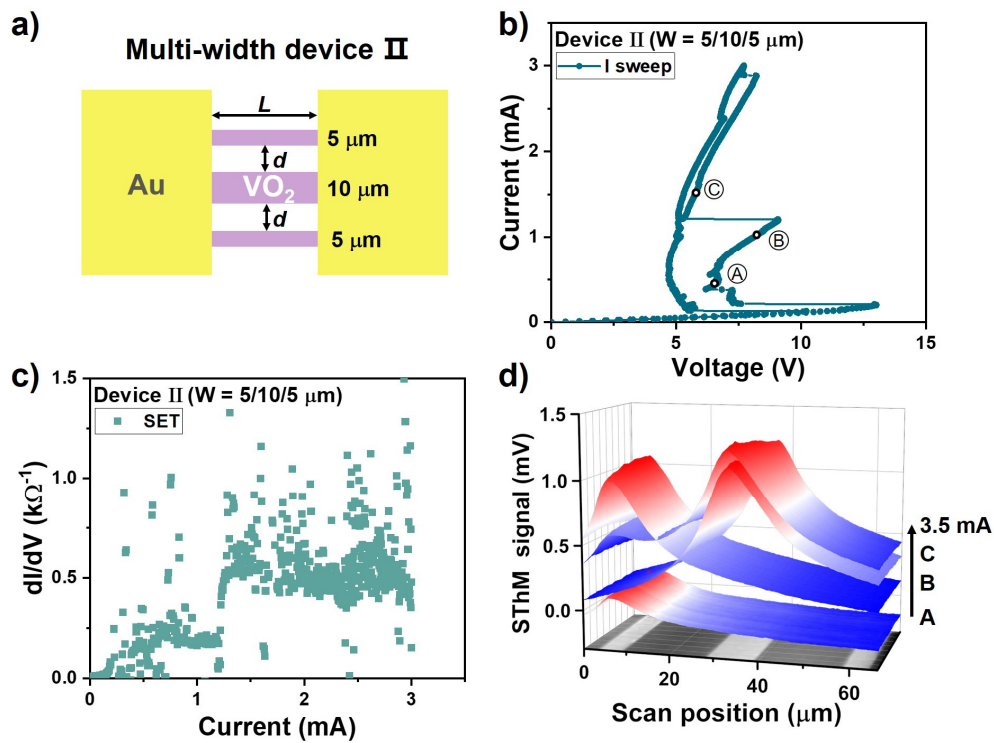

**Figure 6.** a) Schematic of Device II ($L$ = 20 μm, $W$ = 5/10/5 μm, d = 20 μm) (top view). b) Current-controlled *I-V* characteristics and c) *dI/dV* plots of Device II. d) 3D SThM thermal maps during the set process for Device II. The points where SThM measurements were taken are labeled by letters (A: 0.4 mA, B: 1 mA, C: 1.5 mA, and 3.5 mA).

From the examples above, it is clear that there are many possibilities for tuning the device configuration by varying the lengths, widths and spacing of the bridges. For example, a multiple-bridge device containing bridges of different widths, growing from one side to the other, is expected to display a 'waterfall switching' sequence from the narrowest bridge to the widest. The spacing can be adjusted to allow the thermal interaction assisting the subsequent switches instead of interfering with them.

## 4. Conclusion

In conclusion, we have investigated the resistive switching behaviors of $VO_2$-based parallel-bridge devices in a two-terminal configuration. The current-controlled measurements allow a higher degree of control over the resistive states compared to the voltage-controlled ones. The current-controlled switching behaviors are influenced by the intrinsic switching properties of the bridges and the thermal interaction among them. The switching behavior can be manipulated at the device level by adjusting several key factors, including bridge numbers, bridge spacing and bridge geometry, which also interact with each other. With more bridges, there is potential to achieve more switching events and resistive states. The spacing between bridges affects the number of switches and the potential switching bridge. In the limit of small device length and current, stochastic behavior can emerge. The switching principles of single-bridge devices can be extended to multiple-bridge devices, which can act as building blocks for versatile reconfigurable devices. A further degree of freedom that can be introduced to enhance the functionality of such $VO_2$ devices with complex topologies is the incorporation of multiple current/voltage terminals. This will be a topic of further study.

## 5. Experimental Section

*Fabrication of $VO_2$-based parallel-bridge devices*: Epitaxial $VO_2$ thin films with an estimated nominal thickness of 11 nm were deposited on single crystal $TiO_2$ (001) substrates using pulsed laser deposition (PLD) from a polycrystalline $V_2O_3$ target.[8] The distance between the target and sample is ~ 45 mm. A KrF excimer laser ($\lambda$ = 248 nm, 20 ns pulse duration) is used with an energy density of ~ 1.3 J/cm² and a pulse repetition rate of 10 Hz. The growth temperature is 400 ºC and the oxygen background pressure is $10^{-2}$ mbar. After deposition, the samples are cooled at 10 °C/min at the same oxygen pressure. X-ray diffraction (XRD) scans are performed to check the crystalline quality of the film, and atomic force microscopy (AFM) scans are conducted in tapping mode to study the surface topography. Representative XRD patterns and AFM images can be found in the previous work.[20] The as-deposited $VO_2$ films are patterned into parallel bridges with photolithography and $Ar^+$ ion beam etching.[20] Two-terminal devices are fabricated with Ti (4.5 nm)/Au (50 nm) contact pads via RF sputtering and lift-off.

*Electrical measurements*: The switching characteristics of the $VO_2$-based parallel-bridge devices are investigated in a Janis cryogenic probe station with a Keithley 4200A-SCS parameter analyzer applying voltage or current sweeps at room temperature ($T_0 = 295$ K), unless otherwise stated.

*Scanning Thermal Microscopy (SThM)*: SThM is performed using an Asylum AFM and an SThM thermo-resistive probe (Pd on SiN from Bruker). These SThM probes can correlate changes in their electrical resistance with temperature variations in the tip ($R_{probe} \propto T_{probe}$).[35] The SThM probe is electrically connected to an external Wheatstone bridge consisting of two fixed resistances (1 kΩ each), a potentiometer ($R_{pot}$), and the resistance of the probe ($R_{probe}$). SThM measurements are performed in passive mode, with a 0.5 V set point and a 0.5 V tip bias. The potential measured across the bridge ($V_{SThM}$) allows to determine accurately the changes of the electrical resistance of the probe and, hence, temperature variations on the surface of the device. The sample is coated with a 90 nm-thick PMMA-A2 layer for the SThM scans, in order to protect the SThM tip from electrical discharges while biasing the devices. During the SThM scans, the device is biased at constant current bias values when the resistance is settled. The current is ramped up towards a maximum at which the entire device is fully switched to the low-resistance state. Afterwards the current is ramped down back to 0 mA to reset the device. To probe the amount and variation of the thermal background noise, a scan with zero bias current is always performed prior to the first biased SThM scan and after the last. To be noted, the thermal signal measured from SThM cannot be directly related to the surface temperature without certain calibration steps,[28] but serves as a qualitative indication of the currents flowing in the bridges.

# Supporting Information

**Multibridge VO$_2$-Based Resistive Switching Devices in a Two-Terminal Configuration**

*Xing Gao, Thijs J. Roskamp, Timm Swoboda, Carlos M. M. Rosário, Sander Smink, Miguel Muñoz Rojo, Hans Hilgenkamp**

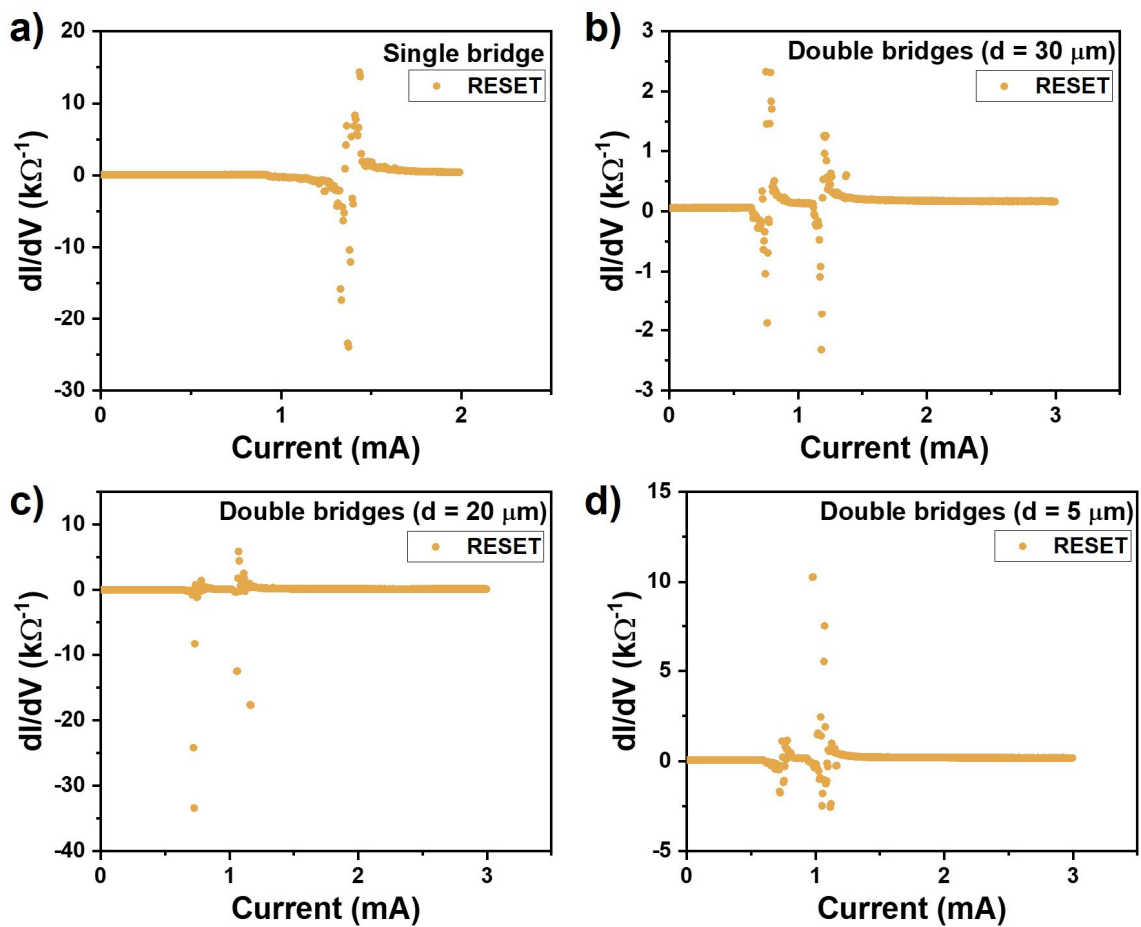

**Figure S1.** Full range of numerical computed differential conductance (*dI/dV*) as a function of applied current during the reset process for the single-bridge and double-bridge devices in a) Figure 1c, b) Figure 1g, (c) Figure 1h, and (d) Figure 1i.

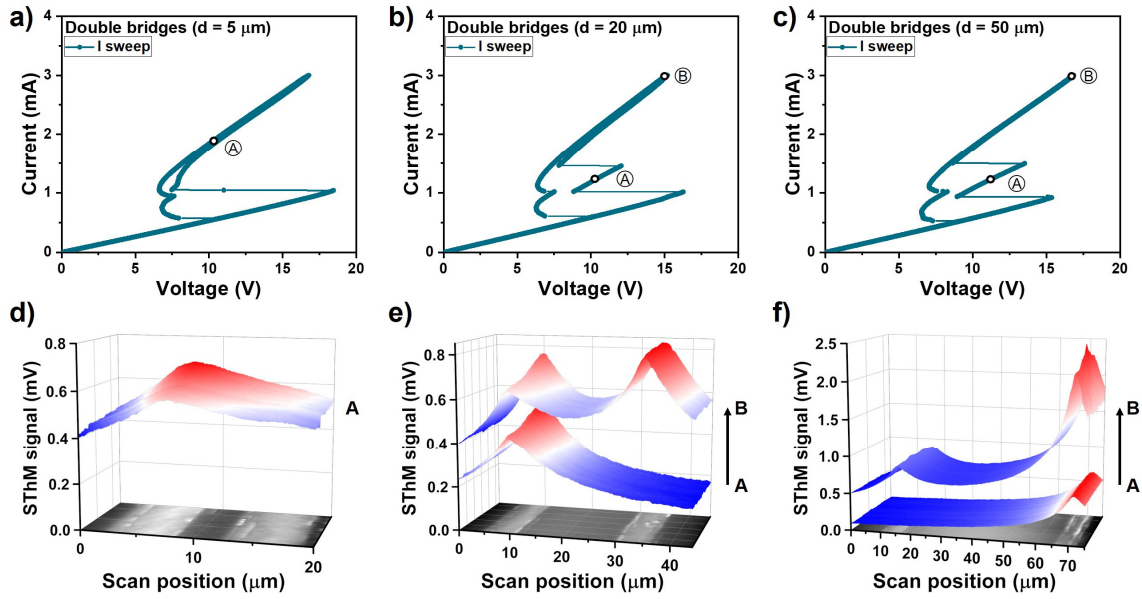

**Figure S2.** Current-controlled *I-V* characteristics of VO$_2$-based double-bridge devices ($L = 20$ μm, $W = 5$ μm) with different spacings: a) $d = 5$ μm, b) $d = 20$ μm, c) $d = 50$ μm. The points where SThM measurements were taken are labeled by A and B, respectively. 3D SThM thermal maps of the corresponding devices: d) $d = 5$ μm, e) $d = 20$ μm, f) $d = 50$ μm. The 2D surface topography images at the bottom are obtained using the SThM tip in Atomic Force Microscopy mode and are shown as a guide for the position of the bridges in the device.

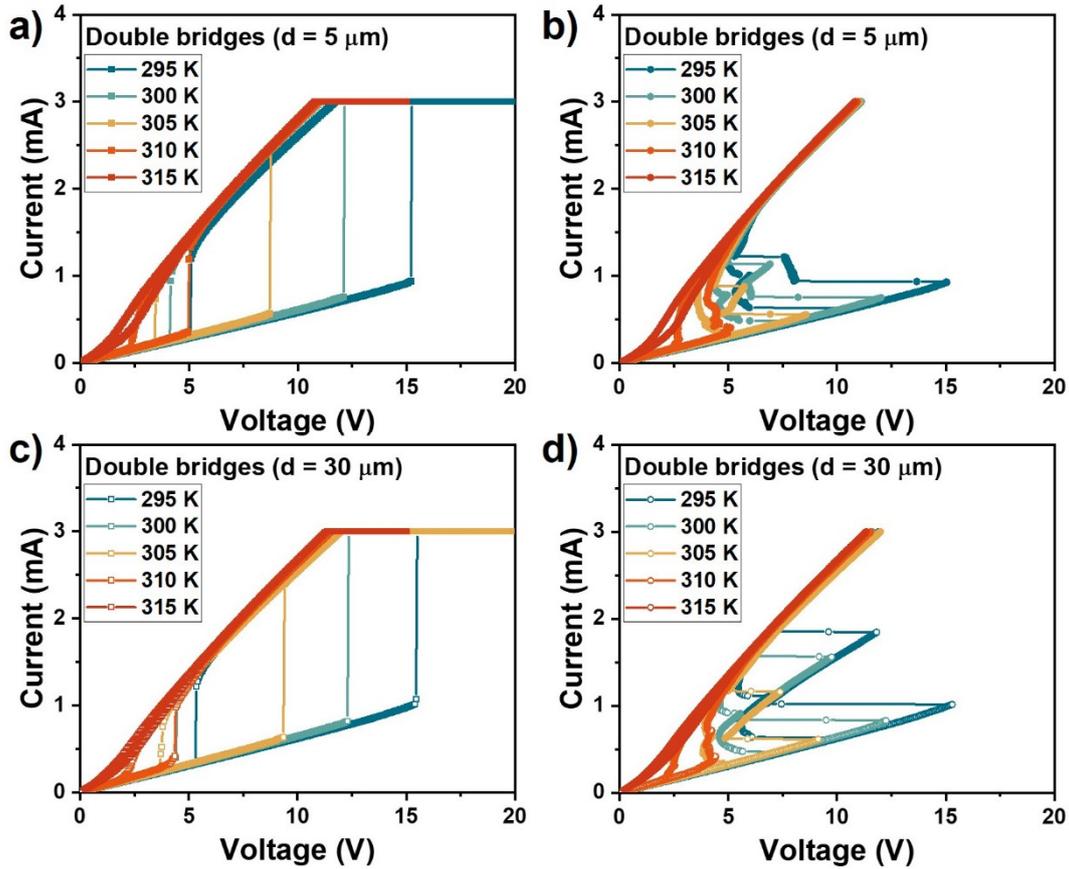

**Figure S3.** Temperature-dependent measurements of VO$_2$-based double-bridge devices ($L$ = 15 μm, $W$ = 5 μm) with different spacings. The temperature increased from 295 K to 315 K with an incremental step of 5 K. a) Voltage-controlled *I-V* characteristics and b) current-controlled *I-V* characteristics for device with a smaller spacing ($d$ = 5 μm). c) Voltage-controlled *I-V* characteristics and d) current-controlled *I-V* characteristics for device with a larger spacing ($d$ = 30 μm).

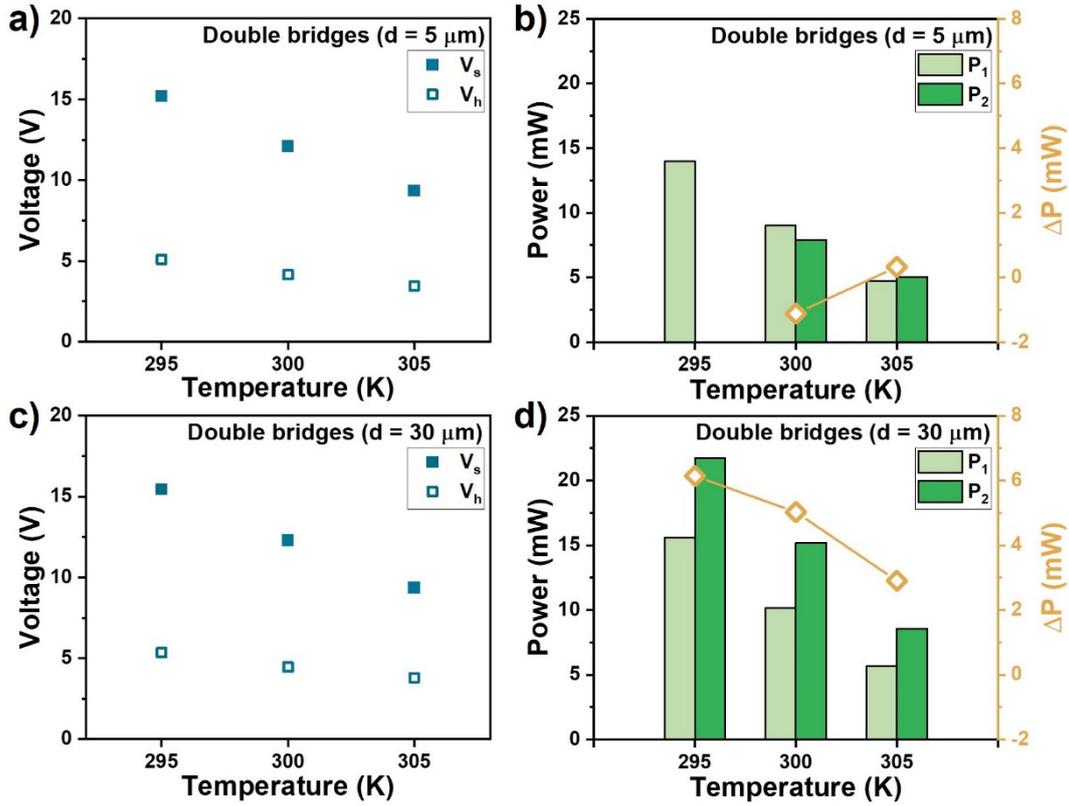

**Figure S4.** Set voltage ($V_s$) and hold voltage ($V_h$) of voltage-controlled measurements as a function of the temperature of $VO_2$-based double-bridge devices ($L$ = 15 μm, $W$ = 5 μm) with different spacings: a) $d$ = 5 μm and c) $d$ = 30 μm. Calculated switching power ($P$) and the power difference ($\Delta P$) between the first and the second snapback of current-controlled measurements as a function of the temperature of $VO_2$-based double-bridge devices ($L$ = 15 μm, $W$ = 5 μm) with different spacing: b) $d$ = 5 μm and d) $d$ = 30 μm.

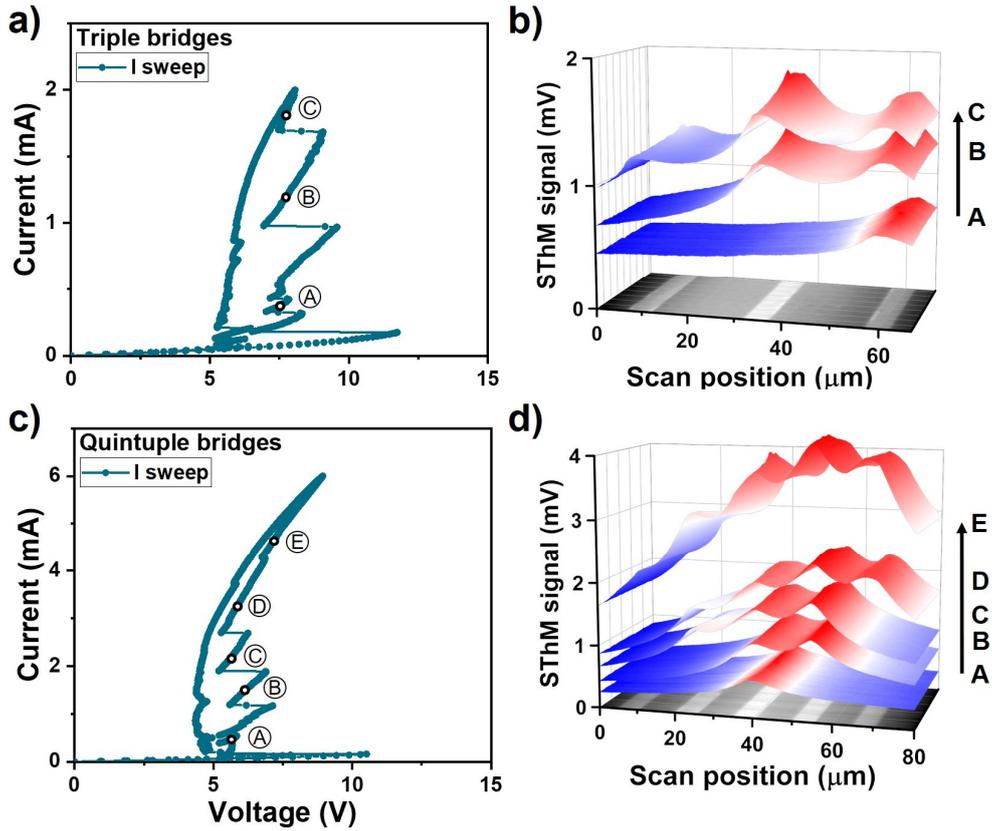

**Figure S5.** a) Current-controlled *I-V* characteristics and b) 3D SThM thermal maps during the set process of a VO$_2$-based triple-bridge devices ($L$ = 20 μm, $W$ = 5 μm, $d$ = 10 μm). The points where SThM measurements were taken are labeled as; A: 0.4 mA, B: 1.2 mA and C: 1.8 mA. c) Current-controlled *I-V* characteristics and d) 3D SThM thermal maps during the set process of a VO$_2$-based quintuple-bridge devices ($L$ = 15 μm, $W$ = 5 μm, $d$ = 10 μm). The points where SThM measurements were taken are labeled by letters (A: 0.4 mA, B: 1.4 mA, C: 2.2 mA, D: 3.3 mA, E: 4.9 mA).

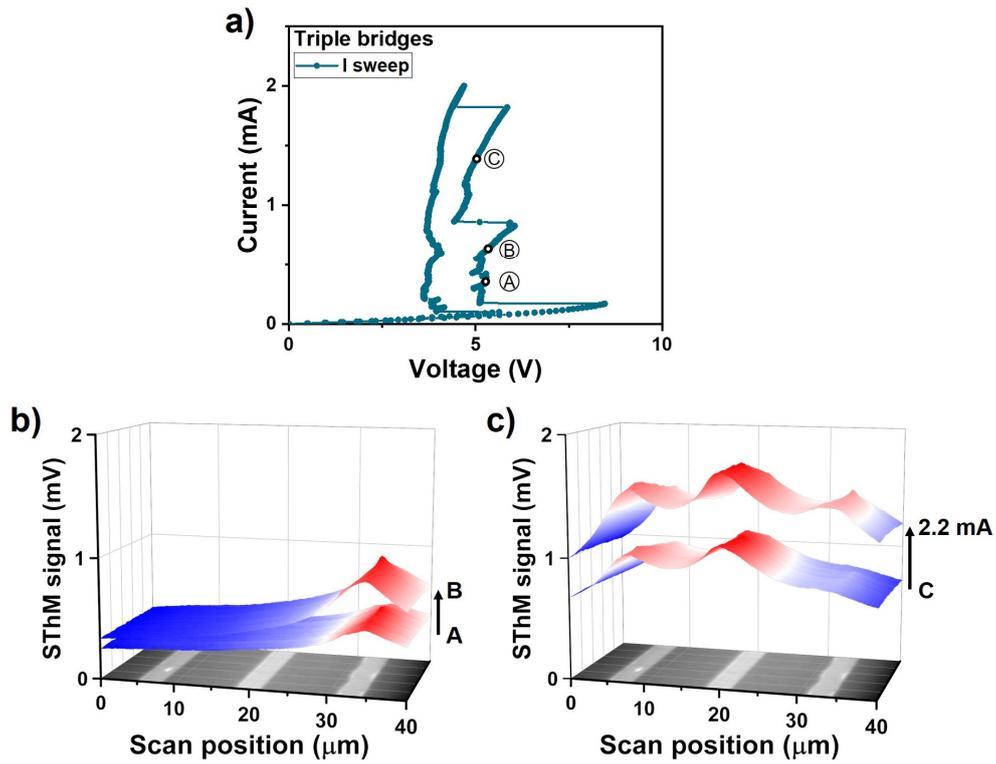

**Figure S6.** a) Current-controlled *I-V* characteristics and b-c) 3D SThM thermal maps during the set process of a VO$_2$-based triple-bridge device ($L$ = 10 μm, $W$ = 3 μm, $d$ = 10 μm). The points where SThM measurements were taken are labeled as; A: 0.35 mA, B: 0.65 mA, C: 1.4 mA, and at 2.2 mA.

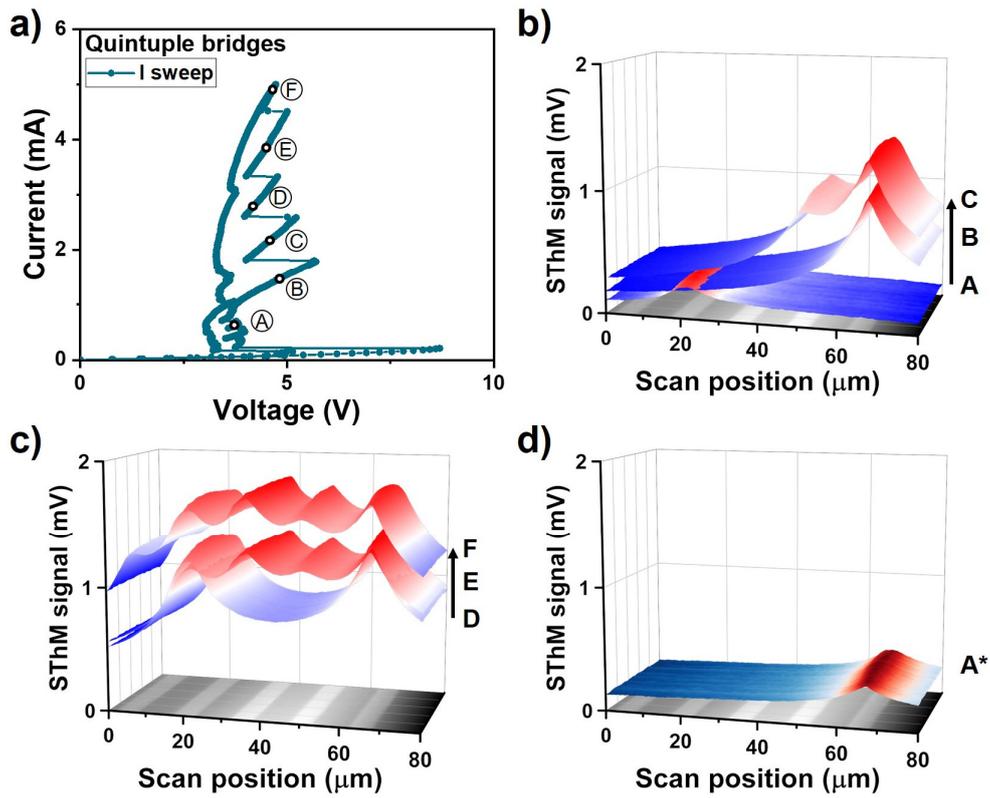

**Figure S7.** a) Current-controlled *I-V* characteristics and b-d) 3D SThM thermal maps during the set process of a VO$_2$-based quintuple-bridge devices ($L$ = 10 μm, $W$ = 5 μm, $d$ = 10 μm). The points where SThM measurements were taken are labeled as; A: 0.4 mA, B: 1.4 mA, C: 2.2 mA, D: 2.8 mA, E: 3.7 mA, and F: 4.9 mA. The SThM scan A* in d) is taken after scan F, at 0.4 mA. A different color scheme is employed in d) to distinguish scan A* from the previous round of SThM scans.